\documentstyle[prb,psfig,aps,epsf,graphics,twocolumn]{revtex}
\begin{document}
\draft
\twocolumn[\hsize\textwidth\columnwidth\hsize\csname
@twocolumnfalse\endcsname
%
%
%

\title{Kinetic energy driven superconductivity \\ and pseugogap phase
in weakly doped antiferromagnets}

\author{P. Wr\'obel$^{1}$, R. Eder$^2$, and R. Micnas$^3$}

\address{$^1$ Institute for Low Temperature and Structure
Research, P. 0. Box 1410, 50-950 Wroc{\l}aw 2, Poland}
\address{$^2$ Institut f\"ur Theoretische Physik, Universit\"at
W\"urzburg,
Am Hubland, 97074 W\"urzburg, Germany}
\address{$^3$ Adam Mickiewicz University, Institute of Physics, Umu{\l}towska
85, 61-614 Pozna\'n, Poland}

\date{\today}
\maketitle

\begin{abstract}
We derive an effective Hamiltonian for spin polarons forming in weakly
doped antiferromagnets and demonstrate that the system becomes
superconducting at finite doping. We argue that the driving mechanism
which gives rise to superconductivity is lowering of the kinetic
energy by formation of mobile antiferromagnetic spin bipolarons. That
source of attraction between holes is by definition effective if the
antiferromagnetic correlation length is longer than the radius of
forming polarons. Notwithstanding that the attraction is strongest in
the undoped system with long range order, the superconducting order
parameter vanishes when the doping parameter decreases which should be
attributed to emptying the spin polaron band and approaching the Mott
insulator phase. Since the hypothetical normal phase of low density
gas of fermions is unstable against formation of bound hole pairs the
intensity of low energy excitations is suppressed and the pseudogap
forms in the underdoped region.
\end{abstract}

\pacs{PACS numbers: 74.20.Mn, 71.10.Fd}

\vskip2pc]
\narrowtext

\section{Introduction}
The appearance of superconductivity (SC) with high $T_c$  in doped
antiferromagnetic (AF)
insulators belongs to  most intriguing problems with which the
contemporary condensed matter physics is confronted. A common
feature of the family of systems which reveals  that phenomenon is
the presence of a building block in the form of copper-oxygen planes.
On the other hand, this class of materials contains many compounds,
properties of which are different in respect of many details. Thus
a task of formulating  a universal model capable of describing
simultaneously all experimental aspects of cuprates seems
elusive. Nevertheless some general understanding of SC in
doped AF may be gained from analysis of a minimal model for such
systems which is the $t$-$J$ model (TJM).

Only recent numerical 
calculations based on combination of various techniques, like
quantum Monte Carlo (QMC) and Lanczos algorithms, performed for
relatively large clusters provided convincing evidence for pairing in
the 
TJM \cite{Sorellaetal02,PoilblancScalapino02}. These calculations also indicate
that short-range AF correlations are robust even for moderate
doping. Some time ago an effective model was suggested to discuss
SC in the TJM \cite{DagottoNazarenkoMoreo95}. According to
that
suggestion the driving attractive force between holes may be
attributed to the fact that by sharing a common link two holes
minimize the the loss of the energy related to breaking  AF links.
This effect was represented in that effective model by a term
corresponding to attraction between holes created at nearest
neighbor sites.  According to a different point of view, pairing in
doped AF is mediated by the exchange of spin waves
\cite{Belinicheretal95,Plakidaetal97}. In this paper we shall
demonstrate that the main energetic gain in the paired state is due to
formation of spin bipolarons which move in a  way that saves
the kinetic energy.   

Detailed knowledge about binding in weakly doped AF, about the role
which symmetry plays in this process \cite{WrobelEder98} and about the
internal structure of the bound pair indicates that the static
attraction between holes related to minimization of the number of
broken AF bonds if a hole pair occupies nearest neighbor sites is
ineffective because in the interesting parameter region $t \gg J$ the
kinetic energy of each hole is raised due to the presence of the
second hole at a nearest neighbor site, which restricts the freedom of
hole motion.  That insight gained by means of the spin polaron
(string) approach based on an assumption that short-range AF
correlations prevail even for moderate doping has been verified by 
extensive comparisons with results of numerical analyses including QMC
\cite{BonisegniManousakis93}, exact diagonalization (ED)
\cite{Dagotto94,Leung02} and density matrix renormalization group
(DMRG) calculations \cite{WhiteScalapino97}. A consistent picture
which emerges from the collection of different pieces of data is that
competition between different phases like the non-superconducting
local pair phase, the SC state, phase separation, or the stripe phase
is governed by an obvious tendency to lower simultaneously the kinetic
and the magnetic exchange energy. That conclusion is only seemingly
trivial, because not like for a weakly correlated system, in which
case the balance between the kinetic and the potential energy takes
more conventional forms, the system has to resort to some tricky ways
to achieve that goal, for example by forming anti-phase AF domains in
the stripe phase \cite{WhiteScalapino98}.

In this paper, using knowledge gained about binding of holes  in
weakly doped AF \cite{WrobelEder98}, we analyze formation of the
SC state in such a system in terms of an effective model, which
represents propagation and interaction of spin polarons. The basic
assumption of this approach is that the AF correlation length is
longer than the radius of spin polarons which seems to be valid at
least in the region of weak doping. We will demonstrate that the shape
of the curve representing the 
superconducting order parameter as a function of doping obtained in
the numerical calculations \cite{Sorellaetal02} is reproduced  within
the Hartree Fock (HF) approximation  to an effective
Hamiltonian represented in the basis of spin polaron states and that the
agreement for underdoped systems where the spin polaron approach
should be valid is satisfactory. 

The standard version of the TJM \cite{Chaoetal78} on the square lattice is used
in
this article, 
\begin{equation}
H=-t \sum_{\langle i,j \rangle, \sigma} \left(\hat{c}^{\dag}_{i,\sigma}
\hat{c }_{j,\sigma}+H.c.\right)+J \sum_{\langle i,j \rangle }
\left( {\bf
S}_i {\bf S}_j -\frac{n_i n_j}{4} \right).
\end{equation}
The $\bf S_i$ are electronic spin operators,
$\hat{c}^{\dag}_{i,\sigma}=c^{\dag}_{i,\sigma} \left( 1- n_{i,-\sigma}
\right)$
and the sum over $\langle i,j \rangle $ stands for a summation over all
pairs of nearest neighbors.

\section{Localized spin polarons and bipolarons}
The spin polaron approach
\cite{Eder92,WrobelEder94,Chernyshevelal94979899} to binding of holes
in doped AF is based on the notion of a string.  A moving hole
inserted into AF medium creates a line of defects (string) in the spin
pattern, which raises the magnetic, potential-like contribution to the
energy. Since the rate of processes related to hopping is higher than
the rate of magnetic exchange processes during which anti-parallel
spins on nearest neighbor sites are turned upside down, the latter
category of processes may be temporarily neglected in the lowest order
approximation, when a trial `unperturbed' Hamiltonian $H_0$ is
solved. That Hamiltonian represents a hole attached to a site by a
string, or in other words it describes a particle in a potential
well. The eigenstates of  the trial Hamiltonian which we call in our
terminology spin polarons span in principle the whole Hilbert space,
but to discuss the low energy properties of the system it is
sufficient to concentrate on the ground-state, which may be
represented as,
\begin{equation}
|\Psi_{ i }\rangle = \sum_{{\cal P}_i} \alpha_{l({\cal P}_i)}
|{\cal P}_i\rangle.
\label{polaron}
\end{equation} 
$|{\cal P}_i\rangle$  denotes a state obtained by
hopping along a path ${\cal P}_i$ without retreats of a hole created at
the site $i$ in the AF medium. For simplicity, we assume that the
N\'eel state plays the role of that medium and that amplitudes
$\alpha_{l({\cal P}_i)}$ depend only on the length of paths $l({\cal
P}_i)$. If more holes are created at distant sites, the wave function
of multi-hole spin polaron representing many holes in separate
potential wells is just a product of wave functions for single
independent polarons. If a hole pair is created at nearest neighbor
sites that approximation can not be applied, because the holes share
the same region in which the spin arrangement has been disturbed. Due
to the size reduction of the disturbed area, the increase of the
static potential contribution to the energy related the part of the
Hamiltonian which is equivalent to the Ising model, is 
reduced. On the other hand, proximity of holes may restrict their
freedom of motion which raises the kinetic energy. In order to
analyze quantitatively these effects we define a localized spin bipolaron
as a combination of states which may be obtained by non-retraceable
hopping of holes created at a pair of nearest neighbor sites $i$, $j$, 
\begin{equation}
|\Psi_{ i,j }\rangle = \sum_{{\cal P}_i,{\cal P}_j} 
\alpha_{l({\cal P}_i),l({\cal P}_j)} |{\cal P}_i,{\cal P}_j
\rangle. 
\label{bipolaron}
\end{equation}
The amplitudes $\alpha_{l({\cal P}_i),l({\cal P}_j)}$ represent the
groundstate solution of an approximate Schr\"odinger equation which
describes two particles in the same potential well. More details
concerning the construction of spin polarons may be found in earlier
papers devoted to analysis of hole binding
\cite{Eder92,WrobelEder94,WrobelEder98}.  A general lesson which we
learn by comparing eigenenergies of localized single polarons and
bipolarons is that the gain in the energy related to the reduction of
the number of broken bonds when holes occupy nearest neighbor sites is
compensated by the loss of the kinetic energy which may be attributed
to the fact that motion of each hole toward its partner is prohibited
in such a case.

\section{Effective Hamiltonian}

\begin{figure}
 \unitlength1cm
\begin{picture}(6.0,8.7)
\epsfxsize=8.3cm
\put(-0.2,0.5){\epsfbox{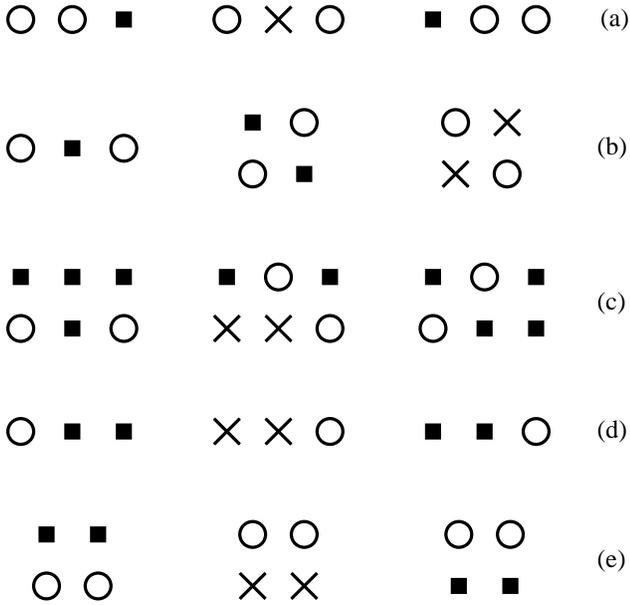}}
\end{picture}
\caption{Schematic representation of some processes contributing to
the Hamiltonian and the overlap operator.}
\label{strings}
\end{figure}

\begin{figure}
 \unitlength1cm
\begin{picture}(6.5,15.5)
\epsfxsize=8.0cm
\put(-0.2,0.5){\epsfbox{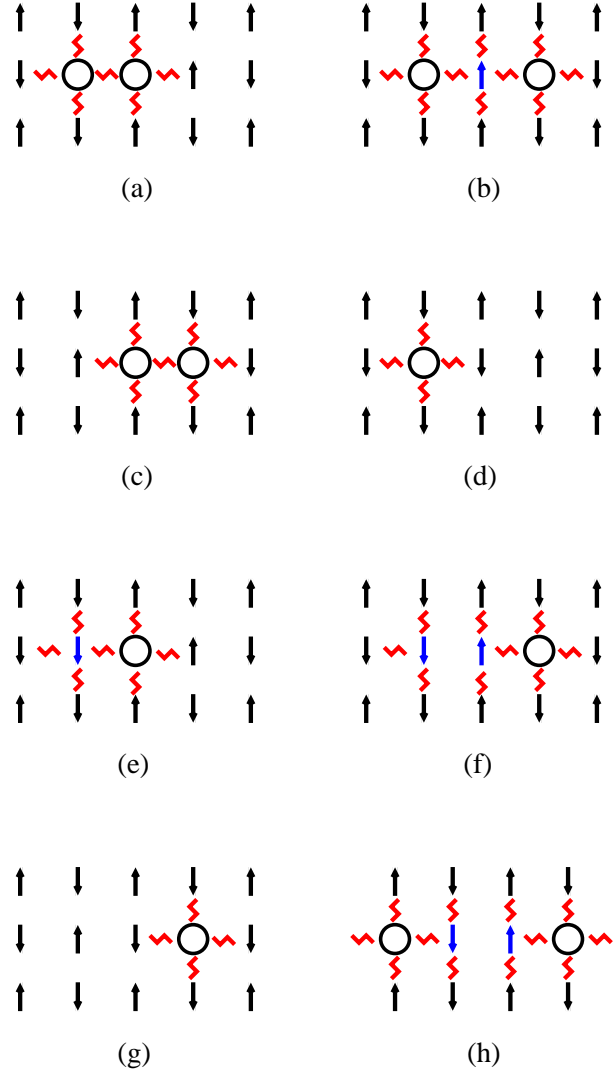}}
\end{picture}
\caption{Explicit representation of some states schematically depicted 
in Fig.\ref{strings}.}
\label{explicit}
\end{figure}

During the process of construction of spin polarons we have solved a
trial `unperturbed' Hamiltonian which is a part of the full TJM. We
will take into account the remaining part of the TJM by analyzing all
processes which have been neglected at the earlier stage of the
calculation. We will express these processes in terms of a Hamiltonian
matrix which couples spin polaron states. This way of expressing the
TJM is very convenient, because eigenenergies of a spin polaron and a
spin bipolaron already contain a substantial part of the energy
related to the fast incoherent motion of holes inside potential
wells. The formulation of the Hamiltonian in terms of the spin-polaron
basis brings about some new features of the formalism. Since spin
bipolaron states are not orthogonal, the particle-number operator
contains two-body terms which consist of a pair of operators
annihilating spin polarons and a pair of operators creating spin
polarons at pairs of nearest neighbor sites. The appearance of such
terms may be understood by means of a simplest example depicted in
Fig.\ref{strings}(a).  A slanted cross in Fig.\ref{strings}(a)
represents a spin turned upside down in comparison with the N\'eel
state, which is a defect in the initial spin background and will be
called a magnon from now onward. A block square denotes a site
occupied by a spin pointing in the same direction as in the initial
N\'eel state.  The left and the right part of Fig.\ref{strings}(a)
represent two holes created in the AF (N\'eel) background on the left
and right pairs of sites, respectively. In order to clarify the
meaning of symbols used in Fig.\ref{strings} we present the same
states in Fig.\ref{explicit}(a)-(c) in an explicit way. A wavy line
represents a frustrated link for which the static contribution to the
exchange energy, which is diagonal in the basis of spin up-down
states, is raised in comparison with the N\'eel state.  The states
depicted in the left and right parts of Fig.\ref{strings}(a) and
Figures \ref{explicit} (a) and (c) are components of two different
bipolarons $ |\Psi_{ i,j }\rangle$ created at two different pairs of
sites $i,j$. In both cases, by hopping outward the accompanying hole,
the hole at the central site creates a state depicted in the middle of
Fig.\ref{strings}(a) and in Fig.\ref{explicit}(b) which is
simultaneously the component of those two different bipolarons.
\noindent 
 Equivalence between components of bipolarons
created at different sites gives rise to the overlap between them. By
further hopping, holes create more equivalent states and the total
overlap between bipolarons created on nearest neighbor sites may be
written as a sum,
\begin{equation}
\omega=-\sum_{\mu=0,\nu=1}(z-1)^{\mu+\nu-1}\alpha_{\mu,\nu}
\alpha_{\mu+1,\nu-1}
,
\end{equation}
 where $z=4$ is the coordination number and the
minus sign is a matter of convention. Analogously, to each string
state of arbitrary length, which consists of aligned magnons and holes
at both end-points, may be attributed overlap between bipolarons
created at outer pairs of sites. In the language of the second
quantization the overlap between bipolarons may be represented in
terms of a pair of  operators annihilating spin polarons and a pair
creating them, as for example,
\begin{equation}
\delta \hat{O}=\omega \sum_{i} 
h^{\dag}_{i+\hat{\bf x}}  h^{\dag}_{i}  h_{i} h_{i-\hat{\bf x}}
\label{do}
\end{equation}
in the previously discussed case, where $\hat{O}$ is an operator
representing the overlap.
\noindent
It turns out that each non-trivial contribution to the overlap
operator brings about a new contribution to the effective
Hamiltonian. By applying the kinetic energy term to the state
represented in the middle of Fig. \ref{strings}(a) and
Fig. \ref{explicit} (b) which is a component of the bipolaron created
at a pair of sites represented by circles in the left part of
Fig.\ref{strings}(a), the left hole may be shifted to the central site
and a state represented by the right part of Fig.\ref{strings}(a) or
Fig.\ref{explicit}(c) will be obtained, which means that spin polarons
created at different pairs of sites marked by circles in outer parts
of Fig.\ref{strings}(a) are coupled by the Hamiltonian. That coupling
was neglected, when we were solving the trial Hamiltonian, because
holes created at a pair of nearest neighbor sites were prohibited to
retrace each other. Longer strings obtained by further hopping of the
right hole may be also involved in analogous processes and the related
contribution to the Hamiltonian is,
\begin{equation}
\delta\hat{H}=-t \sum_{\mu=1}(z-1)^{\mu-1}\alpha_{0,\mu}
\alpha_{0,\mu-1}  \sum_{i} 
h^{\dag}_{i+\hat{\bf x}}  h^{\dag}_{i}  h_{i} h_{i-\hat{\bf x}}.
\label{dh}
\end{equation}
Longer strings are crucial to effectiveness, in lowering the energy, of
processes driven by the kinetic term in the Hamiltonian. They have
been neglected in previous analyses by different authors which lead
them to an incorrect conclusion that the collective motion of a hole
pair connected by a string can not bring about pairing. We will later
discuss that issue.  
At this stage of our considerations it is necessary to mention that
the contributions to the energy which are brought about by the
processes included in the trial Hamiltonian are incorporated into the
eigenenergies of the polaron $E_1$ and the bipolaron $E_2$ and appear
in the effective Hamiltonian as diagonal terms
\begin{equation}
\delta\hat{H}=E_1 \sum_i h^{\dag}_{i}  h_{i}
\end{equation} and 
\begin{equation}
\delta\hat{H}=(E_2 - 2 E_1) \sum_{i,\delta} h^{\dag}_{i+\delta}  h^{\dag}_{i}
h_{i} h_{i+\delta},
\end{equation}
 where $\delta$ denotes links to nearest neighbors of $i$.  Since the
absolute value of amplitudes $\alpha_{\mu}$, $\alpha_{\mu,\nu}$
declines, when the length of strings $\mu$ or $\mu+\nu$ grows, only
short string states may bring about considerable contributions to the
effective Hamiltonian. This remark concerns only the shortest strings
which are involved in a process of a given type as in the
Fig.\ref{strings}(a). As we have mentioned before, also longer strings
obtained by further hopping of holes in the state presented in
Fig.\ref{explicit}(b) may take part in an analogous process. Since the
number of such strings grows exponentially with the length, their
contributions should be also taken into account as we did in
(\ref{do}) and (\ref{dh}). On the other hand, in this paper the
analysis of processes is restricted to shortest strings (of maximal
length of 2 lattice spacings) which are involved in a given process
and `initiate' the whole family of longer strings that may also take
part in it. We apply an obvious defining convention that the string
length in these units is equal to the number of magnons created by 
hopping holes, which means that the maximal distance between holes
connected by a string of length 2 is 3 lattice spacings. Restricting
our calculation to processes which involve strings with the minimal
length not longer than 2 lattice spacings needs some justification. By
solving the Schr\"odinger equation determining the shape of spin
polarons we deduce that the weight of a string state of length 3 is
already smaller at least by one order of magnitude than the weight of
states representing bare holes created in the N\'eel state and drops
faster with the increasing length. Thus, we immediately realize that
the weight of shortest strings involved in a given process basically
determines the order of magnitude of its amplitude which may be also
confirmed by explicit evaluation of formulas like the sums (\ref{do})
and (\ref{dh}). In addition, results of experiments with neutron
scattering performed for $La_{2-x}Sr_xCuO_4$ \cite{Birgenauetal88}
suggest that the AF correlation length in the cuprates follows the the
mean hole distance which allows us to make an estimate that the spin
polaron approach to pairing in weakly doped AF will provide reasonable
results for the doping parameter $\delta \leq 1/9$ for which the AF
correlation length is longer than the average distance between the
holes that form the spin bipolaron, which we estimate to be about 2-3
distances between copper atoms. The applicability of the string
approach to the whole underdoped region, for example for the doping
parameter up to the value $1/4$ starts to be questionable because at
that value the AF correlation length is surely not higher than 2
distances between copper atoms. Thus we may asses the validity of the
spin polaron method for description of pairing for the doping
parameter higher than $\delta \simeq 1/9$ only by comparing with the
results of numerical calculations.

The hypothesis about the tight relation between AF correlations and SC
in the cuprates has been recently confirmed once again by observing
the coexistence of the AF order with the SC state in
$YBa_2Cu_3O_{6.5}$ \cite{Sidisetal01}. That observation suggests that
the region where the spin polaron approach is applicable may indeed be
wider.

In a recent preprint an absence of in plane hole ordering and the
homogeneity of holes in the superconducting compound
$La_2CuO_{4+\delta}$ was observed above $T_c$ by means of anomalous
x-ray scattering at the oxygen K edge \cite{Abbamonteetal02}. On the
other hand the problem of phase separation in the $t$-$J$ model is a
delicate issue. Emery {\it et al.} suggested that the phase separation
in the 2D $t$-$J$ occurs at all interaction strengths
\cite{Emeryetal90}. Results of a high-temperature expansion (HTE)
\cite{Putikaetal92} and of a Green's function Monte Carlo (GMFC)
calculation \cite{Calandraetal98} indicate that the $t$-$J$ model does
not phase separate for the values of the ratio $J/t$ relevant for
HTSC. These conclusions were questioned in a series of papers by
different authors \cite{HellbergManousakis97,HellbergManousakis00} who
found phase separation at half-filling for all values of the
interaction-strength $J/t$. Their findings are little surprising
because in a previous paper \cite{BonisegniManousakis93} they showed
that binding of holes occurs above a certain finite value of the ratio
$J/t$. Our analysis of the 1D $t$-$J$ model in the staggered magnetic
field introduced in order to mimic the physics in 2D suggests that large
clusters of holes are heavy objects and that their formation will mean
a great loss of the kinetic energy which makes the phase separation to
be unlikely at low values of the parameter $J/t$
\cite{WrobelEder96}. In addition, recent DMRG calculations
\cite{WhiteScalapino98,WhiteScalapino00} indicate that the phase
separation occurs in the $t$-$J$ model for $J/t \geq 1.4$ which
roughly agrees with the results of HTE. On the other hand, in the physically
relevant region of parameters stripes are observed in results of DMRG
calculations. Notwithstanding the outcome of the discussion about the
robustness of phase separation in the $t$-$J$ model at half-filling
for the values of the parameter $J/t$ which are relevant from the
physical point of view, it is clear that this phenomenon is very
sensitive to the Coulomb repulsion. This interaction does not spoil
the mechanism of pairing because the binding holes are located at
longer distances. Even if the effects related to Coulomb repulsion are
neglected and even if the statement about the phase separation in the
$t$-$J$ model at half-filling at all interaction strengths is taken
seriously, there seems to exist a region near the doping level
about $\delta \simeq 1/9$ where the system is not phase separated for
physically relevant values of the interaction strength, and the
discussed in this paper formulation of the spin polaron approach is
applicable. On the other hand it is clear that the issue of phase
separation needs further analysis in future, also in the framework of
the string approach. 

Less controversial are experimental \cite{Tranquadaetal95} and
theoretical \cite{WhiteScalapino98,WhiteScalapino00} findings that
stripes form in doped antiferromagnets. The physics of the stripe
phase like the physics of binding is determined by the balance between
the kinetic and the exchange energy. Since the domains between stripes
are antiferromagnetically ordered the methodology of spin polaron
approach may be applied to the analysis of the stripe phase
\cite{Chernyshevetal0002,WrobelEder00}. It is unlikely that the
formation of stripes is a driving mechanism for superconductivity but
pairing may in principle also take place in the stripy background. The
analysis of this phenomenon by means of the spin polaron method may be
done but is beyond the scope of this paper.
 
While constructing the effective Hamiltonian we not only take into
account processes, which were omitted when the trial Hamiltonian was
solved, but also make some amendments to approximations we made
previously. Since the wave function of a spin multi-polaron
representing holes created at a distance longer than one lattice
spacing is approximated by the product of wave functions, some
corrections are necessary. By considering that kind of products we
tacitly assumed that both holes in the left part of Fig.1(b) may
simultaneously jump on the middle site. That artificial state should
be removed, which gives rise to a necessary correction in the
normalization condition for a pair of spin polarons created at sites
marked by empty circles in the left part of Fig.1(b) and an additional
term in the operator representing overlap between pairs of polarons.
Also an appropriate correction to the eigenenergies of single polarons
should be made for polarons created at such a small distance, because
the obvious restriction on possibility of hopping of holes on top of
each other was neglected when we solved the Schr\"odinger equation
defining single polaron states, which means that some spurious
processes were taken into account during the evaluation of the kinetic
energy. There are some more non-existing states which we artificially
incorporated into the calculation, by assuming that the wave function
of spin polarons created at the distance higher than one lattice
spacing may be approximated by a product of wave-functions for
separated polarons. For example, the left hole in the left part of
Fig.\ref{strings}(b) can not move to the right site occupied by the
second hole. Similar restrictions concern two holes and polarons
created at opposite ends of the cell diagonal, as in the middle panel
of Fig.\ref{strings}(b), and should bring about appropriate
corrections both to the operator which represents the overlap between
polarons and to the Hamiltonian formulated in the polaron language. In
addition, the same state in the right panel of Fig.\ref{strings}(b)
may be obtained in two different ways when the upper hole in the
middle part of Fig.\ref{strings}(b) hops once vertically or
horizontally while the lower hole hops in the opposite way, which
gives rise to an additional contribution to the normalization
condition for the wave function of two polarons occupying opposite
corners of an elementary cell.  By means of a detailed analysis of
string states we may find more examples of overlap between pairs of
polarons created at different pairs of sites and more contributions to
the Hamiltonian related to the kinetic energy. By hopping twice, first
to the right and next upward, the left hole in the left part of
Fig.\ref{strings}(c) may transform the state without magnons in the
vicinity of holes into a state represented by the middle part of
Fig.\ref{strings}(c). The same result may be achieved if the left hole
in the right diagram in Fig.\ref{strings}(c) hops twice to the right,
which means that components of two different pairs of polarons
localized at sites represented by circles in outer parts of
Fig.\ref{strings}(c) are identical and that the wave functions of
different pairs of polarons overlap. In addition, a shift of one of
two holes in the middle diagram to the site occupied by the central
magnon, caused by the application of the kinetic term in the TJM
transforms a component of a wave function for one pair of polarons
into the component of another pair, which means that the effective
Hamiltonian couples different pairs of polarons. Until now, we have
considered only two categories of contributions to the Hamiltonian
expressed in the basis of spin polarons, terms which represent
eigenenergies of spin polarons, and terms related to the fast motion
of holes, with the rate $\sim t$, or in other words matrix elements of
the Hamiltonian which represent coupling between spin polarons by the
hopping part of the TJM. Different spin polaron states may be also
coupled by terms in the TJM related to the magnetic exchange. Their
action, which occurs at a slower rate $\sim J$, turns upside down
anti-parallel spins at nearest neighbor sites. That coupling was
neglected when the spin polaron basis was
constructed. Fig.\ref{strings}(d) shows a most obvious process, which
gives rise to coherent propagation of a single hole in the AF
medium. The left diagram represents a hole created in the AF spin
background.  This state is also a component of a spin polaron created
at the left site.  Another component of that polaron depicted in the
middle diagram will be obtained if the hole hops twice from the left
site to the right site. Two magnons created in this way may be
annihilated if the transversal part of the Heisenberg model is applied
to the state represented by the middle part of
Fig.\ref{strings}(d). The new state, which is a component of the spin
polaron localized at the right site, is represented by the right
diagram in Fig.\ref{strings}(d). Four Figures \ref{explicit}(d)-(g)
represent the same sequence: the initial state, two intermediate
states obtained by hopping of the hole and the final configuration in
which two shifted spins has been swapped.  The magnetic exchange may
also couple a spin polaron representing two holes attached by strings
to a pair of nearest neighbor sites with another spin bipolaron or a
pair of separated spin polarons. A process depicted by
Fig.\ref{strings}(e) may be analyzed in a way similar to that which
was applied in the case of the process represented by
Fig.\ref{strings}(d). The intermediate state in the middle of
Fig.\ref{strings}(e) is a component of a spin bipolaron localized at a
lower part of sites marked by circles in the left
diagram. Annihilation of spin defects, or in other words magnons, by
the transversal part of the Heisenberg model gives rise to the
creation of a state represented by the right diagram.  That state is a
component of the spin bipolaron created at the upper pair of sites,
which means that the spin bipolaron is effectively shifted upward by
one lattice spacing. If holes initially created at a lower pair of
sites do not hop in the same direction, an analogous process will
transform the spin bipolaron into a pair of separated spin
polarons. The process depicted in Fig.\ref{strings}(d) is of paramount
importance for the selection of the symmetry of the bound state of two
holes created in an AF \cite{WrobelEder98}, because it lowers the
energy of the $d_{x^2-y^2}$-wave state and raises the energy of the
$p$-wave state, while the rest of low order processes which involve
only spin bipolarons is neutral. The proximity between the energies of
lowest states which show these symmetries has been recently confirmed
by an exact diagonalization performed for a relatively large cluster
\cite{Leung02}. We expect that the preference for the
$d_{x^2-y^2}$-wave symmetry will prevail in the hypothetical SC state,
which may emerge after polaron pairs condense.

Some additional amendments to terms in the Hamiltonian which are
diagonal in the spin-polaron representation are necessary.  For
example, our analysis should also take into account that holes
initially created at sites which are not nearest neighbor sites, may
gain some potential energy by lowering the number of broken bonds,
when they occupy such a pair after they made a few hops. We may meet
such a situation if holes have been initially created at the Manhattan
distance of two lattice spacings like in the middle part of
Fig.\ref{strings}(b). After a hop of a hole toward its companion the
number of broken bonds is lower by one than after a single hop of one
of two holes initially created at a longer distance. The discussion of
quantum fluctuations in the AF state which lower the energy of the
ground state of the Heisenberg model in comparison with the energy of
the N\'eel state that is the groundstate of the Ising model is also
incorporated into our calculation. In the lowest order of the
perturbation theory such fluctuation represent pairs of magnons
created at NN sites in the N\'eel state and change the energy by the
amount $-J/12$ for each link.

At the chosen level of accuracy there
are altogether 14 different contributions to the overlap operator and
58 to the Hamiltonian, which may be classified according to processes
that give rise to them and the positions of involved polarons. The
physical picture which underlies the principle according to which the
Hamiltonian is constructed is based on the assumption that the
dynamics of holes should not destroy local AF correlations.  For
example in the process depicted in Fig.\ref{strings}(a) and Figures
\ref{explicit}(a)-(c) the defects in the spin structure created by the
motion of the right hole are annihilated by the subsequent hopping of
the left hole. Thus, by the exchange of magnons forming a string which
connects two holes, hole pairs initially created at NN sites avoid
confinement. The process depicted in Fig.\ref{strings}(d) and Figures
\ref{explicit}(d)-(g) which deconfines a single hole may be
interpreted as cutting of the string formed by magnons attached to the
initial site, by the transversal part of the exchange term in the
Hamiltonian.

Without dwelling more upon
details we present now the form of the effective Hamiltonian
expressed in terms of operators $h_i$ and $h^{\dag}_i$ 
annihilating and creating spin polarons. 
\begin{eqnarray}
&&\hat{H}-\mu \hat{N}=(E_1-\mu) \sum_{i} h^{\dag}_i h_i +  h_1
\sum_{i,\delta,\delta^{\prime}; \; \delta^{\prime} \neq -\delta}
h^{\dag}_{i+\delta+\delta^{\prime}} h_i \nonumber \\ && + (E_2/2-E_1+u_1)
\sum_{i,\delta} h^{\dag}_{i} h^{\dag}_{i+\delta} h_{i+\delta} h_i
\nonumber \\ &&+ u_2 \sum_{i,\delta,\delta^{\prime}; \; \delta^{\prime} \neq
-\delta} h^{\dag}_{i} h^{\dag}_{i+\delta+\delta^{\prime}}
h_{i+\delta+\delta^{\prime}} h_i \nonumber \\ && + u_3
\sum_{i,\delta,\delta^{\prime}; \; \delta^{\prime} \perp \delta}
h^{\dag}_{i} h^{\dag}_{i+\delta+\delta^{\prime}}
h_{i+\delta+\delta^{\prime}} h_i \nonumber \\ && + u_4
\sum_{i,\delta,\delta^{\prime},\delta^{\prime \prime}; \;
\delta^{\prime} \neq - \delta, \delta^{\prime \prime} \neq
-
\delta^{\prime}} h^{\dag}_{i}
h^{\dag}_{i+\delta+\delta^{\prime}+\delta^{\prime \prime}}
h_{i+\delta+\delta^{\prime}+\delta^{\prime \prime}} h_i \nonumber
\\ && + s_1
\sum_{i,\delta,\delta^{\prime}; \; \delta^{\prime} \neq -\delta}
h^{\dag}_{i+\delta+\delta^{\prime}} h^{\dag}_{i+\delta}
h_{i+\delta} h_i \nonumber \\ &&+ s_2
\sum_{i,\delta,\delta^{\prime},\delta^{\prime \prime}; \;
\delta^{\prime} \neq -\delta, \delta^{\prime \prime} \neq
-\delta^{\prime}} h^{\dag}_{i+\delta+\delta^{\prime}}
h^{\dag}_{i+\delta+\delta^{\prime}+\delta^{\prime \prime}}
h_{i+\delta} h_i \nonumber
\\ &&  + s_3 \sum_{i,\delta,\delta^{\prime};
\; \delta^{\prime} \perp \delta} [(h^{\dag}_{i}
h^{\dag}_{i+\delta+\delta^{\prime}} h_{i+2 \delta} h_i +H.c.)\nonumber \\ &&+
h^{\dag}_{i} h^{\dag}_{i+\delta+\delta^{\prime}} h_{i+
\delta-\delta^{\prime}} h_i] \nonumber
\\ && + s_4
\sum_{i,\delta,\delta^{\prime},\delta^{\prime \prime}; \;
\delta^{\prime} \neq - \delta, \delta^{\prime \prime} \neq
-
\delta^{\prime}} (h^{\dag}_{i}
h^{\dag}_{i+\delta+\delta^{\prime}+\delta^{\prime \prime}}
h_{i+\delta} h_i + H.c.) \nonumber \\ &&+ s_5 \sum_{i,\delta,\delta^{\prime};
\;
\delta^{\prime} \perp \delta} h^{\dag}_{i}
h^{\dag}_{i+\delta^{\prime}} h_{i+\delta} h_i  \nonumber \\ &&  +
s_6 \sum_{i,\delta,\delta^{\prime},\delta^{\prime \prime}; \;
\delta^{\prime} \neq \delta, \delta^{\prime \prime} \neq
-
\delta} (h^{\dag}_{i+\delta+\delta^{\prime \prime}}
h^{\dag}_{i+\delta^{\prime}} h_{i+\delta} h_i + H.c.) \nonumber \\ &&+ s_7
\sum_{i,\delta,\delta^{\prime}; \; \delta^{\prime} \perp \delta}
h^{\dag}_{i+\delta+\delta^{\prime}} h^{\dag}_{i+\delta^{\prime}}
 h_{i+\delta} h_i
\label{ham}
\end{eqnarray}
Parameters which appear in this effective Hamiltonian are functions of
$E_1$, $E_2$, $\mu$, $t$, $J$ and amplitudes $\alpha$ and include at
once contributions from many different types of processes. An
important remark which we should also make is that the highest value
for experimentally relevant ratios $J/t$ has a parameter related to
the motion as a whole of strings connecting a pair of holes, example
of which is depicted in Fig \ref{strings}(a). That type of
caterpillar-like motion is so effective in lowering of the total energy
because by expanding at one end and shrinking at the other, the whole
string may move freely, while the number of magnetic defects is kept
low. Only the kinetic term in the Hamiltonian is involved in that
movement and the term related to the magnetic exchange does not have
to intervene. Thus the gain in the energy is mainly due to lowering of
the kinetic energy. 

It is widely believed \cite{Trugman98,Carlsonetal98,Carlsonetal02}
that the motion of the hole pair is frustrated and can not bring about
lowering of the total energy and binding or pairing.  Already the
analysis of hole binding \cite{WrobelEder98} provided arguments that
such an opinion is not correct. The notion of frustration was used in
literature to describe the fact that effective hopping of the hole
pair occupying NN sites, to nearest links which are parallel and
perpendicular to the link at ends of which the holes have been
initially located, produces effective hopping integrals with the same
positive sign, which is not very convenient in terms of lowering the
kinetic energy, but does not change a generally applying rule that a
mobile quantum object has lower energy than an immobile one. We have
previously shown, that the motion of a hole pair connected by a string
formed by defects in the AF spin structure may give rise to formation
of bound states with $d_{x^2-y^2}$ and $p$-wave symmetries, which
agrees with results of numerical analyses including a recent work
\cite{Leung02} performed for a relative large cluster consisting of 32
sites. Also the energetic hierarchy of two-hole states representing
symmetries and wave vectors allowed by the geometry of the $4\times 4$
cluster observed by Hasegawa and Poilblanc \cite{HasegawaPoilblanc89}
in the results of the exact diagonalization has been reproduced by
means of the spin polaron approach.  Since the interaction between
spin polarons mediated by the processes related to the motion of the
string connecting two holes is dominating, the agreement between
numerical and analytical analyses indicates that the spin polaron
approach properly takes into account such effects. Arguments against
the kinetic energy driven mechanism of binding in doped AF are based
on the large $d$ expansion \cite{Carlsonetal02}.  A single hole
created in the N\'eel background may lower the energy by virtual
hopping to NN sites. If two holes occupy NN sites, the hopping of each
hole in one direction is blocked and the energy is raised by the
amount $2t^2/Jd$ in comparison with the energy of two separate
holes. On the other hand if holes are created at NN sites one spoiled
AF link is saved and a negative contribution $-J/2$ to the total
energy is generated. In the first order of the $1/d$ expansion, the
propagation of a hole pair occupying NN sites mediated by the process
represented by Fig.\ref{strings}(a) may only compensate the loss in
the energy related to the blocking effect and no net gain in the
energy related to the hole-pair kinetics is observed. That picture
changes qualitatively in lower dimensions for $t\gg J$.  The energy
scale $\sim t^2/J(d-1)$ dominates the scale $\sim J(d-1)$ and the energetical
cost related to creation of longer strings similar to the state
represented by Fig.\ref{explicit}(h) is relatively lower. In addition,
there is no blocking effect in the case of strings with at least one
magnon.  In simple words, holes at ends of longer strings can hop at
least once in all directions without disturbing each other.  All this
makes the creep of strings more effective in lowering the energy.  In
2D any simple calculation based on the $1/d$ expansion or an approach
limited to a small basis of states related to short strings will not
provide reliable results. During the construction of the spin polaron
and the bipolaron the important contribution to the energy from
incoherent motion in the potential wells and longer strings has also
been taken into account.  The energy of the spin bipolaron by
construction contains contributions related to saving spoiled AF links
and mutual restricting the freedom of motion by two holes which
oscillate chaotically around a pair of NN sites where they have been
initially created. It turns out that for physically relevant range of
parameters these two effects almost compensate each other and the
eigenenergy of the localized spin bipolaron is roughly twice the
energy of a localized polaron. Thus, truly kinetic effects related to
motion of the center of mass of a hole pair connected by a string
bring about a net gain in the kinetic and total energy.  In the
effective Hamiltonian these effects are represented by terms related
to hopping of bipolarons.  The difference between behavior in low and
high dimensions may be associated to the change in the relation
between the energy scales $J(d-1)$ and $t^2/J(d-1)$ and to the related fact
that the creation of longer strings is not so costly in lower
dimensions.

If phase separation indeed takes place at half filling
the Coulomb repulsion between holes at NN sites may prevent that
effect, but will not influence pairing so much, because the total
probability that holes which form a spin bipolaron are at a longer
distance is much higher than they occupy NN sites. A detailed analysis
of effects related to the Coulomb repulsion is beyond the scope of
this paper.

The wave functions of spin polarons are not orthonormal and  the operator
$\hat{O}$  representing overlap between them takes an unconventional form,
\begin{eqnarray}
\hat{O}&=& 1+d_1 \sum_{i,\delta,\delta^{\prime}; \; \delta^{\prime}
\neq -\delta} h^{\dag}_{i} h^{\dag}_{i+\delta+\delta^{\prime}}
h_{i+\delta+\delta^{\prime}} h_i \nonumber \\ &&  + d_2
\sum_{i,\delta,\delta^{\prime}; \; \delta^{\prime} \perp \delta}
h^{\dag}_{i} h^{\dag}_{i+\delta+\delta^{\prime}}
h_{i+\delta+\delta^{\prime}} h_i   \nonumber
\\ && + o_1
\sum_{i,\delta,\delta^{\prime}; \; \delta^{\prime} \neq -\delta}
h^{\dag}_{i+\delta+\delta^{\prime}} h^{\dag}_{i+\delta}
h_{i+\delta} h_i  \nonumber \\ && + o_2
\sum_{i,\delta,\delta^{\prime},\delta^{\prime \prime}; \;
\delta^{\prime} \neq -\delta, \delta^{\prime \prime} \neq
-\delta^{\prime}} h^{\dag}_{i+\delta+\delta^{\prime}}
h^{\dag}_{i+\delta+\delta^{\prime}+\delta^{\prime \prime}}
h_{i+\delta} h_i \nonumber
\\ &&  + o_3 \sum_{i,\delta,\delta^{\prime};
\; \delta^{\prime} \perp \delta} [(h^{\dag}_{i}
h^{\dag}_{i+\delta+\delta^{\prime}} h_{i+2 \delta} h_i +H.c.)  
\nonumber \\ && +
h^{\dag}_{i} h^{\dag}_{i+\delta+\delta^{\prime}} h_{i+
\delta-\delta^{\prime}} h_i].
\end{eqnarray}
Since the explicit formulas for the parameters of the Hamiltonian and
the overlap operator are rather lengthy, they will be presented in
the Appendix. 
                
\section{Pairing versus pseudogap phase in doped antiferromagnets}
The distance between two holes which form a bound state in the
AF background is few lattice spacings \cite{WrobelEder98} and it is
natural to analyze their paring in the real space. That approach is
suitable for superconductors with the short coherence length. For the
sake of
simplicity we concentrate on anomalous Green's functions
${\cal F}(i,\tau;i^\prime,\tau^\prime)$, 
\begin{equation}
{\cal F}(i,\tau;i^\prime,\tau^\prime)=
\langle T_{\tau} h_i(\tau) h_{i^\prime}(\tau^\prime) \rangle,
\end{equation}
which represent a pair of spin polarons simultaneously annihilated at
a pair of sites $i$, $i^\prime$ located at the distance not longer
than 3 lattice spacings. The rest of anomalous Green's functions which
corresponds to longer distances is neglected. That simplification will
be justified by showing that ${\cal F}(i,\tau;i^\prime,\tau^\prime)$
decreases fast with the distance between $i$ and $i^\prime$. Our
intention is to reproduce the results of the recent numerical
calculation performed by Sorella and collaborators
\cite{Sorellaetal02} by means of numerical methods. Since they observe
pairing correlations at some pairs of nearby sites we may also define
the order parameter in the real space for a few short distances. Since
attraction between holes is strongest in the $d_{x^2-y^2}$-wave
channel we shall for simplicity discuss pairing only of that
symmetry. Possible symmetries of the order parameter are determined by
irreducible representations of the point group $C_{4v}$. The order
parameter which is a singlet may transform according to one
dimensional representations $s$, $d_{x^2-y^2}$, $d_{xy}$ and $g$,
while the triplet order parameter corresponds to the two dimensional
representation $p$. The pairing at the distance of 1, 2 or 3 lattice
spacings may in principle realize the symmetries $s$, $d_{x^2-y^2}$ and
$p$. Scattering of a hole pair mediated by the process represented by
Fig.\ref{strings}(a) and similar caterpillar-like motion of longer
strings analogous to the object depicted in Fig.\ref{explicit}(h),
constitutes the strongest interaction in the Hamiltonian, which
determines the dominating symmetry of pairing. This property was
already noticed when we discussed binding \cite{WrobelEder98}. That
kind of interaction is attractive in the $d_{x^2-y^2}$ and $p$-wave
channels and repulsive in the $s$-wave channel. Another term in the
Hamiltonian related to the process depicted in Fig.\ref{strings}(e)
favors $d_{x^2-y^2}$ and suppresses $p$-wave pairing. Therefore, the
dominating terms in the order parameter will have $d_{x^2-y^2}$
symmetry. Interactions which involve holes at slightly longer
distances are also relevant. If we restrict pairing in the real space
to distances up to 3 lattice spacings, the $d_{x^2-y^2}$ symmetry will
generate in the order parameter three independent harmonics
$D^{(1,0)}_{\bf k}$, $D^{(2,1)}_{\bf k}$ and $D^{(3,0)}_{\bf k}$
defined in the Appendix. Necessity to apply a multi-dimensional
non-monotonic order parameter was recently suggested after analysis of
results of some experiments with Raman scattering performed for
electron doped cuprates \cite{Blumbergetal02}. The order parameter
representing pairing of holes at the distance $\sqrt{2}$ may transform
according to representations $s$, $d_{xy}$ and $p$, while for the
paring at the distance $\sqrt{5}$ according to the representations
$s$, $d_{x^2-y^2}$, $d_{xy}$, $g$ and $p$. In a full analysis of
pairing in the real space at distances not longer than 3 lattice
spacings we should in principle consider a 24 dimensional order
parameter, which is much beyond the scope of this paper, and we will
concentrate only on the dominating pairing in the $d_{x^2-y^2}$-wave
channel.

We assume that the anomalous Green's function is translationally
invariant in space and time, 
\begin{equation}  
{\cal F}_e({\bf x},\tau)=\langle T_{\tau} h_{i+{\bf
x}}(\tau^\prime+\tau) h_{i}(\tau^\prime)\rangle, 
\end{equation}
and that $i$ in the previous definition belongs to the even
sublattice.
A relevant order parameter in the real space
is defined as, 
\begin{equation}
\Delta_{\bf x} ={\cal F}_e({\bf x},0^+). 
\end{equation}
 By proceeding in a standard way we derive HF equations for
the SC order parameter in which  vertex corrections have been
neglected. Since 
retardation effects related to the exchange of magnons have been
already taken into account during the derivation of the effective
Hamiltonian, the application
of the weak coupling approach seems to be appropriate. For the sake of
brevity we 
shall present only in the Appendix the difference between the grand
canonical potential in the superconducting and the normal state which
may be reconstructed from the equations for the order parameter. The
chemical potential applied in this formalism refers to the number of
holes $\hat{N}$ which is given by the  formula, 
\begin{equation}
\hat{N}=\sum_{i} h^{\dag}_i h_i + 2 (\hat{O}-1),
\nonumber
\end{equation}  
that within the HF approximation may be written at $T=0$ as 
\begin{equation}
\delta=\frac{1}{N}
\sum_{\bf k}(1-\frac{\xi_{\bf k}}{E_{\bf k}})/2
-8(o_1-o_2)\Delta_{\hat{\bf e}_x}^2,
\label{hnsc} 
\end{equation}
where 
\begin{equation} 
\delta=\langle \hat{N} \rangle /N.
\end{equation}
Quasiparticle energies $\xi_{\bf k}$ and  $E_{\bf k}$ in the normal
and SC state are defined in the Appendix. Since we deal with a
two-dimensional (2D) system where fluctuations are strong at finite
temperatures and destroy the long range order we perform the analysis
only at $T=0$.   

\begin{figure}
 \unitlength1cm
\begin{picture}(6.0,4.8)
\epsfxsize=12.2cm
\put(-0.3,-7.5){\epsfbox{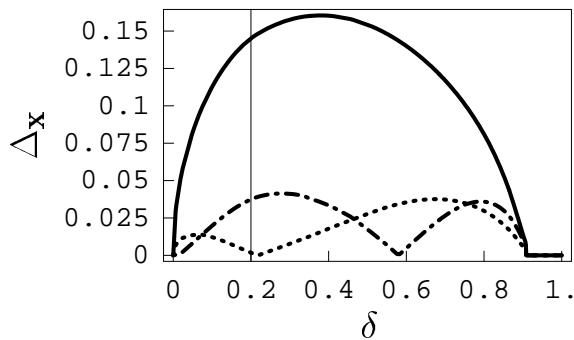}}
\end{picture}
\caption{Anomalous Green's functions $\Delta_{\bf x}$  which represent
pairs of spin 
polarons condensing at the distance of $1$ lattice spacing (continuos
line), $\sqrt{5}$ lattice spacings (dotted line) and  $3$ lattice
spacings (dash-dotted line).}
\label{deltas}
\end{figure}

$o_1-o_2$ turns out
to  be negative. Thus it is clear that Eq.\ref{hnsc} will enforce
disappearance of the SC order parameter when the number of holes
decreases.  
Fig.\ref{deltas} depicts anomalous Greens function that represent the SC
order parameter related to pairs of spin polarons condensing on some
nearest pairs of sites in the real space obtained within the weak
coupling approximation applied to the effective
Hamiltonian for $J/t=0.33$.

Numerical analyses of the TJM indicate that the AF correlations
decrease with doping and that the correlation length becomes
comparable to two or even one lattice spacing, when the doping exceeds
$20 \%$ which agrees with the phenomenology of the cuprates. Since the
spin-polaron approach is based on the assumption that the polaron
radius is smaller than the correlation length, we can not expect that
our analysis will give any reasonable solutions to the problem of
pairing in doped AF for the values of the doping parameter $\delta$
right to the vertical line drawn in Fig.\ref{deltas} at $\delta=0.2$.
On the other hand the agreement for the underdoped system between the
analytical approach and numerical results of Sorella and collaborators
\cite{Sorellaetal02} is reasonable at the left side of the vertical
line which exceeds our earlier expectations that the string approach
should provide correct results for doping levels $\delta$ below
$1/9$. Robustness of the SC solution up to the doping level $\delta
=0.9$ indicates that the disappearance of SC in the overdoped region
should not be attributed merely to the change in the band-filling, but
to the reconstruction of the effective interaction which mediates
pairing. The low energy Hamiltonian for spin polarons which we use is
not filling-dependent and does not take into account effects related
to the degradation of spin polarons at higher doping levels.
Our calculation also demonstrates that the energetic gain is not due
to minimization of the number of broken bonds in the AF state if holes
reside on a pair of nearest neighbor sites, but pairing actually
occurs because by formation of spin bipolarons, the magnetic and
kinetic components of the energy may be simultaneously
lowered. Since $t \gg J$, lowering of the kinetic energy plays a
leading role in pairing, which confirms recent experimental
observations \cite{Molengraafetal02} that the spectral weight in the plots of
optical conductivity is shifted toward lower energies below
$T_c$ and in the pseudogap region. We also observe that the SC
order parameter vanishes in the limit of low hole doping, which may be
attributed to emptying the spin polaron band and approaching the Mott
insulator (MI) phase in the nominally half-filled system.  A rigid
band picture which may be associated with propagating spin polarons
has been recently observed by means of ARPES measurements in the
Na-doped $Ca_2Cu0_2Cl_2$ by Shen, Takagi and collaborators
\cite{Kohsakaetal01,Damascellietal02}. The only effect which has
doping up to the level above $10\%$ is the shift of the chemical
potential. In addition, some recent measurements of optical
conductivity in underdoped cuprates performed by Basov and
collaborators \cite{Basovetal02} indicate that approaching the
insulator regime in this system may be attributed to localization
effects in an band which is emptied.  

The vanishing of the SC energy gap related to the coherent SC state
does not necessarily mean that the underdoped AF should reveal
features of an ordinary Fermi liquid-like normal state, because the
system of freely propagating spin polarons becomes unstable against
formation of bipolarons which brings about opening of the a pseudogap
in the spectral function of a single quasiparticle. An earlier
analysis \cite{WrobelEder94} together with results presented in this
paper and some numerical calculations \cite{Dagotto94,Leung02}
demonstrate that the binding energy of a hole pair in the AF medium,
which according to our scenario should be the energy scale of the
pseudogap near half-filling is a bigger fraction of J than the SC
energy gap at optimal doping, which is in rough agreement with the
phenomenology of the cuprates. 

There exists a direct relation between formation of bound hole-pairs
of $d_{x^2-y^2}$-wave symmetry and existence of current-current
correlations that form the pattern of a staggered flux
\cite{Ivanovetal00,Leung00,WrobelEder01}, which makes a connection
between the local pair scenario for the pseudogap phase and the idea
of a hidden order in the $d$-density wave phase
\cite{Chakravartyetal01}. Some earlier analyses also suggest that the
the appearance of the stripe phase in weakly doped AF is very likely,
because gains achieved by formation of stripes and spin polarons
\cite{Chernyshevetal0002} or bipolarons \cite{WrobelEderII00} are
similar. Localization effects which should accompany a transition to
MHI in the limit of low doping and the intrinsic disorder
\cite{ChenSchrieffer02} which is a characteristic feature of the
cuprates may also additionally complicate the physical picture of the
relation between the superconducting state and the pseudogap phase.

\section{Conclusions}
\begin{figure}
 \unitlength1cm
\begin{picture}(6.0,1.0)
\epsfxsize=7.3cm
\put(0.2,0.3){\epsfbox{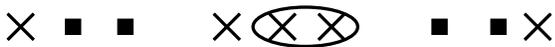}}
\end{picture}
\caption{A process responsible for spin wave propagation.}
\label{sw}
\end{figure}

The analysis presented in this paper allows to identify spin
fluctuations which mediate pairing.  It turns out that the coherent
propagation of magnons which takes the form of spin waves is not
relevant to pairing. A process which gives rise to magnon propagation
has been depicted in Fig.\ref{sw}. 
\noindent A magnon at the left site in the left panel will be
effectively shifted to the right site in the right panel if the
transverse part of the Heisenberg model is applied to the state
depicted in the middle panel which represents the magnon and an
additional spin fluctuation, included in an oval, that exists in the
ground state of the 2D AF. Inclusion of such processes into
considerations does not influence the results, which proves that the
standard propagation of spin waves does not play a crucial role in
pairing.  Emission of spin fluctuation by a hopping hole, which are
later annihilated by the second hole that retraces the first hole has
a quite different nature and is a dominating factor in pairing.  On
the other hand, the coupling between the spin degrees of freedom
represented by independently propagating magnons, and the charge which
may be attributed to holes, gives rise to unconventional excitations
in doped AF for example with the charge $Q=1$ and spin $S_z=3/2$ which
have been observed in the numerical calculation by Hasegawa and
Poilblanc \cite{HasegawaPoilblanc89}. Such a quasiparticle may be
interpreted as a bound state of the excess magnon and the hole. It is
possible to analyze that effect in the framework of the spin polaron
approach. However, interaction between independently propagating spin
waves and holes does not influence the properties of the paired state
and that kind of coupling has been neglected in this paper.  A general
statement that pairing in the region of small doping is driven by spin
fluctuations is rather an oversimplification. Formation of spin
polarons and bipolarons in the locally AF background is actually a
complex process of lowering simultaneously the exchange energy and the
kinetic energy.

The
form of the effective Hamiltonian (\ref{ham}) indicates that the
pairing mechanism is essentially nonretarded and effective interactions
have a short range which agrees with  conclusions drawn from the
universal trends observed for the cuprates in the dependence of $T_c$
on the hole and condensate density \cite{Uemuraetal1991}.     

According to our scenario, phase fluctuations may play a lesser role
in formation of the pseudogap than it was earlier suggested
\cite{EmeryKivelson95} because even in the mean field approximation
the order parameter for SC vanishes in the limit of weak doping, which
follows directly from the relation between the chemical potential,
number of holes and the SC order parameter. It is, however, clear that
fluctuation effects are visible in all extreme type II superconductors
with a short coherence length and that they play an essential role in
the thermodynamic behavior. For example, the measured thermodynamic
properties of cuprate superconductors reveal consistency with the
critical behavior of the 3D $XY$ model \cite{SchneiderKeller}. On the
other hand the disproportion between the vanishing gap function in the
limit of low doping and the finite energy of the hole-pair binding is
more fundamental to the spectral properties of a single quasiparticle
and the formation of the pseudogap. Notwithstanding that the
Hamiltonian (\ref{ham}) is exclusively written in terms of fermionic
operators it is clear that it will reveal some features of the local
pair physics \cite{Micnasetal90}. According to some earlier
estimations \cite{WrobelEder98}, in the case of two holes created in
the quantum AF about $70 \%$ of the weight in the $d_{x^2-y^2}$-wave
bound state belongs to the bipolaron wave function which represents
two holes connected by strings consisting of magnons. Those strings
are pinned to pairs of nearest neighbor sites, which determine the
position of bipolarons. It is sufficient to consider only hopping of
bipolarons, mediated by shrinking and expanding of strings at opposite
ends, to understand the energetical hierarchy of two-hole states that
posses $s$, $d_{x^2-y^2}$ and $p$ symmetries, which indicates that the
local pair physics may be relevant in the limit of low doping. On the
other hand, a more appropriate way of thinking about the spin polaron
Hamiltonian would be from the point of view of a more general
boson-fermion model (BFM) \cite{Robaszkiewiczetal87}. The
identification of spin bipolarons with bosons may serve as
justification of the BFM, which is widely analyzed in the context of
the pseudogap formation. There exist differences between the standard
BFM and a version of it which corresponds to the spin polaron
Hamiltonian (\ref{ham}) but the basic behavior should be similar in
both cases. The main differences are that spin bipolarons (hard-core
bosons) occupy links and that sites at ends of such a link can not be
occupied by fermions. Fermions (mono-polarons) at nearby but not
nearest neighbor sites transform into bosons when they move to NN
sites and vice versa. The energy of the boson localized in space is
not much different from the energy of two localized fermions and this
energy difference is not crucial to pairing. A driving mechanism of
effective attraction between fermions and formation of bipolarons
(bosons) is the high mobility of the latter. Condensing spin
bipolarons predominately contribute to the SC state of a short
coherence length.

We can draw an interesting conclusion concerning the pseudogap in
doped AF from the comparison of the left panel in Fig.\ref{deltas}
with the seemingly useful right panel. If AF spin fluctuations are
responsible for pairing, which we believe is generally true, the mechanism of
attraction between quasiparticles will become less effective when
doping increases and our model becomes irrelevant right to the
vertical line in Fig.\ref{deltas}.  Thus, the fast decrease of $T_c$
with doping observed in real systems and the fast disappearance of the
order parameter demonstrated in the numerical analysis should be
attributed to diluting of the spin system an disappearance of the driving
factor which is identified
as AF short range correlations. That remark concerns also the
temperature of the crossover
to the pseudogap behavior, which decreases with doping already in the
underdoped region and somewhere near optimal doping merges with
$T_c$. The lowering of $T_c$ in the underdoped region with the
decreasing number of holes should not be attributed to the
disappearance of the attractive force between quasiparticles but to
emptying  the spin polaron band. The later effect does not influence
binding of hole pairs which gives rise to pseudogap phenomena. That
rough scenario will in reality be modified by dimensionality effects,
phase fluctuations, and some other phenomena like tendency toward
phase separation and stripe formation. A more detailed analysis of
pseudogap formation in the spin polaron model for weakly doped AF in
the framework of a method suitable for low density systems, which is
the T matrix approach \cite{Kadanoffetal} is beyond the scope of this
paper. 

A phenomenological scenario in which superconductivity is
mediated by lowering of the kinetic energy has been suggested by Hirsch
\cite{Hirsch00a,Hirsch00b}, who introduced the notion of hole
undressing. It seems that the same term may be applied to describe
physics of spin polaron pairing. Single holes form mono-polarons which
are heavy objects because their propagation is mediated by the process
represented by Fig.\ref{strings}(d) which involves the action of the
exchange term and the effective hopping amplitude is therefore
small. Two bound holes may propagate collectively without intervention
of the exchange interaction and they behave more like bare undressed
holes. This remark agrees with experimental observations that HTSC
cuprates reveal transition to a more coherent state as the system
becomes SC. Microscopic models analyzed by Hirsch and their physics
are quite different from our effective Hamiltonian and the spin
polaron scenario. The analysis of these models lead to a conclusion
that the undressing scenario may apply only to holes. On the contrary,
at the level of the $t$-$J$ model, the spin polaron scenario is also
applicable to electron doped systems in which the physics of doubly
occupied sites (particles) is the same as the physics of holes in hole
doped systems.

The analysis of pairing in weakly doped AF in terms of bipolaron
formation seems to be     
complementary to the spin fluctuation approach \cite{Chubukovetal02}
formulated by Pines and collaborators for optimally doped and
overdoped systems.    

In summary, by constructing an effective Hamiltonian we have
identified spin fluctuations which mediate pairing in doped AF as
local spin fluctuations which lie on a path connecting two
holes. Creep-like motion of the whole object is an effective way of
lowering the kinetic energy and the predominant factor which gives
rise to pairing.  This contradicts previous statements and widespread
opinions based on the $1/d$ expansion that the collective motion of
two holes in the locally AF background can not effectively lower the
energy.  The experimental evidence that pairing may be associated with
the change in the kinetic energy has been recently found by a second
group \cite{Santander-Syroetal02}.  These researchers were actually
looking for a transfer of spectral weight from lower to higher
energies in the pseudogap region, but observed an opposite behavior,
which supports the suggestion that the physics of the pseudogap may be
also related to binding of holes.

The analysis of the effective Hamiltonian obtained by means of a method which
has both a variational and perturbative character reproduces the behavior of
the SC order parameter obtained by means of numerical analysis
\cite{Sorellaetal02} with reasonable accuracy up to the doping level $\delta
\approx 0.2$, which exceeds our expectations that the spin polaron method
should be valid for $\delta \leq 0.11$. We also provided  arguments that even
within a most favorable scenario for phase separation, the pairing consistent
with the string picture will occur in the region around the doping level
$\delta = 1/9$. Data presented in a recent review \cite{Sorella02} of numerical
approaches to strongly correlated lattice models indicate that pairing
correlations are smooth functions of the Manhattan distance which indicates
that the physics of strings is visible in these results. Some recent
experimental results obtained by measuring optical conductivity
\cite{Basovetal02} and ARPES \cite{Kohsakaetal01} confirm the relevance of the
band scenario at low doping which is consistent with the spin polaron approach.      

The strength of the
attraction mediated
by spin fluctuations decreases with doping and the diminishing AF
correlation length. That effect may explain decreasing of the
pseudogap with doping and the disappearance of superconductivity in the
overdoped  region. On the other hand in the underdoped region SC
disappears, because the quasiparticle band is emptied, while the density
of low energy excitations is still suppressed by formation of
bound hole pairs. 

\acknowledgements
One of the authors (P.W.) acknowledges support by the Polish Science
Committee
(KBN) under contract No. 5 P03B 058 20. He also highly appreciates
hospitality of KOSEF, Han-Yong Choi and his collaborators from the  Sung
Kyun Kwan University in Suwon South Korea, where a part of this work was
performed. 
\vspace{0.5cm}

\appendix
We start the Appendix with the presentation of parameters which define
the Hamiltonian and the 
overlap operator in terms of operators creating and annihilating
spin polarons 
\begin{eqnarray}
h&=&2M_{\{ (2,0) \} \{ (0,0) \}} \\
 u_1&=&J^H_{\{ {(0,0) \atop (1,0) } \} \{  {(0,0) \atop (1,0) } \} }-
J^H_{\{ {(0,0) \atop (3,0) } \} \{  {(0,0) \atop (3,0) } \} }
 \\ 
u_2&=&(E_1-\mu)P_{\{ {(0,0) \atop (2,0) } \} \{  {(0,0) \atop (2,0) } \}\}} +
P^H_{\{ {(0,0) \atop (2,0) } \} \{  {(0,0) \atop (2,0) } \} } /2
 \nonumber \\
&&+J^H_{\{ {(0,0) \atop (2,0) } \} \{  {(0,0) \atop (2,0) } \} } \\
 u_3&=&(E_1-\mu)R_{\{ {(0,0) \atop (1,1) } \} \{  {(0,0) \atop (1,1) } \} }+
R^H_{\{ {(0,0) \atop (1,1) } \} \{  {(0,0) \atop (1,1) } \} }\\
u_4&=&J^H_{\{ {(0,0) \atop (3,0) } \} \{  {(0,0) \atop (3,0) } \} }/2 \\ 
s_1&=&(E_2- 2 \mu) 
C_{\{ {(2,0) \atop (1,0) } \} \{  {(0,0) \atop (1,0) } \} }-
2M_{\{ (2,0) \} \{ (0,0) \}} \nonumber \\
&& + C^H_{\{ {(2,0) \atop (1,0) } \} \{  {(0,0) \atop (1,0) } \} }\\
s_2&=&(E_2- 2 \mu) 
C_{\{ {(2,0) \atop (3,0) } \} \{  {(0,0) \atop (1,0) } \} }
+ C^H_{\{ {(2,0) \atop (3,0) } \} \{  {(0,0) \atop (1,0) } \} } \\
s_3&=& (2 E_1 - 2 \mu -J/2) 
S_{\{ {(0,0) \atop (1,1) } \} \{  {(0,0) \atop (2,0) } \} } +
S^H_{\{ {(0,0) \atop (1,1) } \} \{  {(0,0) \atop (2,0) } \} } \\
s_4&=& M_{\{ {(0,0) \atop (3,0) } \} \{  {(0,0) \atop (1,0) } \} }+
M_{\{ {(0,0) \atop (1,0) } \} \{  {(0,0) \atop (3,0) } \} } \nonumber \\
&&-2 M_{\{ (2,0) \} \{ (0,0) \}}  \\
s_5&=& 2 M_{\{ {(0,0) \atop (0,1) } \} \{  {(0,0) \atop (1,0) } \} }-
2 M_{\{ {(0,0) \atop (3,0) } \} \{  {(0,0) \atop (1,0) } \} } \nonumber \\
&& -2 M_{\{ {(0,0) \atop (1,0) } \} \{  {(0,0) \atop (3,0) } \} } +
2M_{\{ (2,0) \} \{ (0,0) \}}  \\
s_6&=&  M_{\{ {(2,0) \atop (-1,0) } \} \{  {(0,0) \atop (1,0) } \}}/2 \\
s_7&=& 2 M_{\{ {(1,1) \atop (0,1) } \} \{  {(0,0) \atop (1,0) } \} }-
  M_{\{ {(2,0) \atop (-1,0) } \} \{  {(0,0) \atop (1,0) } \} }
 \nonumber \\
&&-(J/2) C_{\{ {(2,0) \atop (3,0) } \} \{  {(0,0) \atop (1,0) } \} }\\
d_1&=& P_{\{ {(0,0) \atop (2,0) } \} \{  {(0,0) \atop (2,0) } \} }/2\\
d_2&=& R_{\{ {(0,0) \atop (1,1) } \} \{  {(0,0) \atop (1,1) } \} }/2\\
o_1&=& C_{\{ {(2,0) \atop (1,0) } \} \{  {(0,0) \atop (1,0) } \} }\\
o_2&=& P_{\{ {(2,0) \atop (3,0) } \} \{  {(0,0) \atop (1,0) } \} }\\
o_3&=& S_{\{ {(0,0) \atop (1,1) } \} \{  {(0,0) \atop (2,0) } \} }
\end{eqnarray}
Parameters which are  presented below correspond to different
categories of process which involve string states.
\begin{eqnarray}
&&P_{\{ {(0,0) \atop (2,0) } \} \{  {(0,0) \atop (2,0) } \} }=
-\Huge[ 2 \sum_{\mu=2,\nu=0} (z-1)^{\mu+\nu-2} \alpha^2_\mu
\alpha^2_{\nu} \nonumber \\
&&+ \sum_{\mu=1,\nu=1} (z-1)^{\mu+\nu-2} \alpha^2_\mu \alpha^2_{\nu}
\Huge] \\&&
C_{\{ {(2,0) \atop (1,0) } \} \{  {(0,0) \atop (1,0) } \} }=
-  \sum_{\mu=0,\nu=1} (z-1)^{\mu+\nu-1} \alpha_{\mu,\nu}
\alpha_{\mu+1,\nu-1} \\&&
C_{\{ {(2,0) \atop (3,0) } \} \{  {(0,0) \atop (1,0) } \} }=
\sum_{\mu=0,\nu=2} (z-1)^{\mu+\nu-2} \alpha_{\mu,\nu}
\alpha_{\mu+2,\nu-2} \\&&
S_{\{ {(0,0) \atop (1,1) } \} \{  {(0,0) \atop (2,0) } \} }=
-\sum_{\mu=2,\nu=0} (z-1)^{\mu+\nu-2} \alpha_{\mu}  \alpha_{\nu} 
\alpha_{\mu-2}   \alpha_{\nu+2} \\ &&
R_{\{ {(0,0) \atop (1,1) } \} \{  {(0,0) \atop (1,1) } \} }=
-\left[ \alpha^2_1+(z-2)\sum_{\mu=2} (z-1)^{\mu-2} \alpha^2_{\mu}
\right]^2\\&&
M_{\{ (2,0) \} \{ (0,0) \}}=(J/2) \sum_{\mu=2} (z-1)^{\mu-2}
\alpha_{\mu} \alpha_{\mu-2} 
\\&&
M_{\{ {(0,0) \atop (3,0) } \} \{  {(0,0) \atop (1,0) } \} }= (J/2)
\sum_{\mu=0,\nu=2} (z-1)^{\mu+\nu-2} \times \nonumber 
\\&&
 \alpha_{\mu,\nu} \alpha_{\mu} \alpha_{\nu-2} 
\\&&
M_{\{ {(0,0) \atop (1,0) } \} \{  {(0,0) \atop (3,0) } \} }= (J/2)
\sum_{\mu=2,\nu=0}\big[\delta_{\mu,2}(z-1)^\nu \nonumber \\ &&+(1-
\delta_{\mu,2})(z-2)(z-1)^{\mu+\nu-3}\big] 
\alpha_{\mu} \alpha_{\nu}  \alpha_{\mu-2,\nu}\\ 
\\&&
M_{\{ {(0,0) \atop (0,1) } \} \{  {(0,0) \atop (1,0) } \} }= (J/2)
\sum_{\mu=2,\nu=0}\big[\delta_{\mu,2} \nonumber +(1-
\delta_{\mu,2}) \times \\ &&(z-2)(z-1)^{\mu-3}\big] \big[ \delta_{\nu,0}
\nonumber
+(1-
\delta_{\nu,0}) (z-2)(z-1)^{\nu-1}\big] \times\\ && 
\alpha_{\mu,\nu}  \alpha_{\mu-2,\nu}\\ 
\\&&
M_{\{ {(2,0) \atop (-1,0) } \} \{  {(0,0) \atop (1,0) } \} }= (-J/2)
\sum_{\mu=1,\nu=1} 
(z-1)^{\mu+\nu-2}  \alpha_{\mu,\nu} \times\\ &&
\alpha_{\mu-1}\alpha_{\nu-1}\\ &&   
M_{\{ {(1,1) \atop (0,1) } \} \{  {(0,0) \atop (1,0) } \} }= (-J/2)
\sum_{\mu=1,\nu=1}\big[\delta_{\mu,1} \nonumber +(1-
\delta_{\mu,1}) \times \\ && (z-2)(z-1)^{\mu-2}\big] \big[ \delta_{\nu,1}
\nonumber
+(1-
\delta_{\nu,1}) (z-2)(z-1)^{\nu-2}\big] \times\\ && 
\alpha_{\mu,\nu}  \alpha_{\mu-1,\nu-1} 
\\&&
C^H_{\{ {(2,0) \atop (1,0) } \} \{  {(0,0) \atop (1,0) } \} }= -t
\sum_{\mu=1} (z-1)^{\mu-1} \alpha_{0,\mu}\alpha_{0,\mu-1}\\&&
C^H_{\{ {(2,0) \atop (3,0) } \} \{  {(0,0) \atop (1,0) } \} }= t
\sum_{\mu=2} (z-1)^{\mu-2}
\alpha_{0,\mu}\alpha_{1,\mu-2} \\&&
P^H_{\{ {(0,0) \atop (2,0) } \} \{  {(0,0) \atop (2,0) } \} }= -2t (
\alpha_{0}^2\alpha_{1}\alpha_{2}+\alpha_{0}\alpha^3_{1}) 
\\&&
S^H_{\{ {(0,0) \atop (1,1) } \} \{  {(0,0) \atop (2,0) } \} }= -t
\sum_{\mu=2}
\alpha_{\mu}\alpha_{0}\alpha_{\mu-2}\alpha_{1}\\&&
R^H_{\{ {(0,0) \atop (1,1) } \} \{  {(0,0) \atop (1,1) } \} }= -t
\sum_{\mu=1}
\alpha_{1}\alpha_{\mu}^2\alpha_{0}\\ &&   
J^H_{\{ {(0,0) \atop (1,0) } \} \{  {(0,0) \atop (1,0) } \} }= (-J/2)
\big[2 \sum_{\mu=2,\nu=0}(z-1)^{\mu+\nu-2} \alpha_{\mu,\nu}^2+\\&&
\sum_{\mu=1,\nu=1}(z-1)^{\mu+\nu-2} \alpha_{\mu,\nu}^2\big] \\ &&   
J^H_{\{ {(0,0) \atop (3,0) } \} \{  {(0,0) \atop (3,0) } \} }= (-J/2)
\big[2 \sum_{\mu=2,\nu=0}(z-1)^{\mu+\nu-2}\times \\&&
(\alpha_{\mu}\alpha_{\nu})^2+
\sum_{\mu=1,\nu=1}(z-1)^{\mu+\nu-2} (\alpha_{\mu}\alpha_{\nu})^2\big]\\ &&   
J^H_{\{ {(0,0) \atop (2,0) } \} \{  {(0,0) \atop (2,0) } \} }= (-J/2)
 \sum_{\mu=1,\nu=0}\big[\delta_{\mu,1}+  \nonumber \\ &&
(1-\delta_{\mu,1})(z-2)(z-1)^{\mu-2}
\big] \big[\delta_{\nu,0}+ \nonumber \\ && 
(1-\delta_{\nu,0})(z-2)(z-1)^{\nu-1}
\big](\alpha_{\mu}\alpha_{\nu})^2.
\end{eqnarray}
Spin polarons are defined as a solution of the following eigenvalue
problem
\begin{eqnarray}
&&z  t \alpha_1 + 2 J \alpha_0 = E_1
\alpha_0 \nonumber, \\ &&
 t \alpha_{\mu-1}+(z-1) t \alpha_{\mu+1}+J\left(\frac{5}{2}
+\mu \right)\alpha_\mu= \nonumber \\ && E_1\alpha_\mu,
\end{eqnarray}
where $\mu \ge 1$. A solution to the following Schr\"odinger equation 
for two particles in the same potential well determines the the wave
function of the spin bipolaron,
\begin{eqnarray}
&& t \left[\alpha_{\mu-1,\nu} + (z-1)  \alpha_{\mu+1,\nu} +
\alpha_{\mu,\nu-1} + (z-1)  \alpha_{\mu,\nu+1}\right]+ \nonumber \\ &&
J\left(4+\mu+\nu-\frac{1}{2}\delta_{\mu+\nu,0}\right)
\alpha_{\mu,\nu}=E_2
\alpha_{\mu,\nu},
\end{eqnarray}
where $\alpha_{\mu,\nu}=0$ for $\mu<0$ or $\nu<0$.
The normalization conditions for spin-polaron wave functions are 
\begin{eqnarray}
\alpha_0^2+ z\sum_{\mu=1}(z-1)^{(\mu-1)}{\alpha}^2_{\mu}=1
\\
\sum_{\mu=0,  \nu=0}(z-1)^{(\mu+\nu)}{\alpha}^2_{\mu,\nu}=1.
\end{eqnarray}
The thermodynamic potential of the model at 
$T=0$  in the approximation applied in the paper is given by 
\begin{eqnarray}
&&\frac{\Omega_s-\Omega_n}{N}|_{T=0}=\frac{1}{N}\sum_{\bf
k}\frac{|\xi_{\bf k}|-\epsilon_{\bf k}}{2} - \nonumber \\&& \{ \Delta_{1,0}^2(
4
u_1+8 u_4 - 4 s_1 + 4 s_2 - 16 s_4 - 8 s_5 + 16 s_6 + 8 s_7)
\nonumber\\ &&+ \Delta_{2,1}^2 24 u_4 +\Delta_{3,0}^2 4 u_4 +
\Delta_{1,0} \Delta_{2,1} (16 s_4 + 16 s_6) + \nonumber \\&& \Delta_{1,0}
\Delta_{3,0} (8 s_4 + 8 s_6) \},
\end{eqnarray}
where $\xi_{\bf k}$ and $\epsilon_{\bf k}$ are  quasiparticle energies
 in the normal and superconducting state 
 \begin{eqnarray}
\xi_{\bf k}&=& E_1 + h (S^{(2,0)}_{\bf k}+ 2 S^{(2,0)}_{\bf
k})-\mu, \\ E_{\bf k} &=&\sqrt{\xi_{\bf k}^2+\Delta_{\bf
k}^2}. \nonumber
 \end{eqnarray}
The gap function is strongly anisotropic,
\begin{eqnarray}
&&\Delta_{\bf k}=d^{(1,0)}_{\bf k} \Delta_{{\bf e}_x}
+d^{(2,1)}_{\bf k} \Delta_{2{\bf e}_x+{\bf e}_y} + d^{(3,0)}_{\bf
k} \Delta_{3{\bf e}_x} \\ && d^{(1,0)}_{\bf k}=( 2 u_1+4
u_4 - 2 s_1 + 2 s_2 - 8 s_4 - 4 s_5 + \nonumber \\ && 8 s_6 + 4 s_7)
D^{(1,0)}_{\bf k}+( 2 s_4 + 2 s_6)D^{(2,1)}_{\bf k}+ \nonumber \\ && ( 2 s_4 +
2
s_6)D^{(3,0)}_{\bf k} \\&& d^{(2,1)}_{\bf k}=( 4 s_4 + 4
s_6)D^{(1,0)}_{\bf k}+6 u_4 D^{(2,1)}_{\bf k} \\ &&
 d^{(3,0)}_{\bf k}=( 2 s_4 + 2
s_6)D^{(1,0)}_{\bf k}+2 u_4 D^{(3,0)}_{\bf k}, 
 \end{eqnarray}
 where
 \begin{eqnarray}
&&D^{(1,0)}_{\bf k} =2 \cos (k_x) - 2 \cos(k_y) \\
&&D^{(2,1)}_{\bf k} =2 \cos (2 k_x+k_y) +2 \cos (2 k_x-k_y)- \nonumber 
\\ &&2
\cos ( k_x+2 k_y) -2 \cos ( k_x-2 k_y) \\&& D^{(3,0)}_{\bf
k} =2 \cos ( 3 k_x) - 2 \cos(3 k_y)  \\ && S^{(2,0)}_{\bf
k} =2 \cos (2 k_x) + 2 \cos(2 k_y) \\ && S^{(1,1)}_{\bf k}
=2 \cos ( k_x+k_y) +2 \cos ( k_x-k_y).
 \end{eqnarray}

\end{document}